\documentclass[lettersize,journal]{IEEEtran}
\usepackage[utf8]{inputenc}
\usepackage{xcolor}
\usepackage{array}
\usepackage{float}
\usepackage{textcomp}
\usepackage{multirow}
\usepackage{amsmath}
\usepackage{graphicx}
\PassOptionsToPackage{normalem}{ulem}
\usepackage{ulem}
\usepackage{subscript}

\makeatletter

\DeclareTextSymbolDefault{\textquotedbl}{T1}
\providecommand{\tabularnewline}{\\}
\providecolor{lyxadded}{rgb}{0,0,1}
\providecolor{lyxdeleted}{rgb}{1,0,0}

\DeclareRobustCommand{\lyxsout}[1]{\ifx\\#1\else\sout{#1}\fi}

\let\oldforeign@language\foreign@language
\DeclareRobustCommand{\foreign@language}[1]{%
  \lowercase{\oldforeign@language{#1}}}

\usepackage{amsfonts}\usepackage{algorithmic}
\usepackage{algorithm}
\usepackage{array}
\usepackage{stfloats}
\usepackage{url}
\usepackage{verbatim}
\usepackage{cite}
\@ifundefined{showcaptionsetup}{}{%
 \PassOptionsToPackage{caption=false}{subfig}}
\usepackage{subfig}
\makeatother

\begin{document}
\markboth{}{Berdasco \MakeLowercase{\textit{et al.}}: A Sample Article Using
IEEEtran.cls for IEEE Journals}
\title{AMC-backed Twin Arrow Antenna for Wearable Electronic Travel Aid System
at 24 GHz}
\author{A. Flórez Berdasco \IEEEauthorblockA{\textit{}},
M. E. de Cos Gómez \IEEEauthorblockA{\textit{}}, J. Laviada,
F. Las-Heras \IEEEauthorblockA{\textit{}} 
\thanks{}\thanks{This work was supported in part by the Ministerio de Ciencia e Innovación
of Spain under the Formación Personal Investigador (FPI) Grant MCIU-20-PRE2019-089912
and under Project META-IMAGER PID2021-122697OB-I00 and in part by
the Gobierno del Principado de Asturias under Project AYUD-2021-51706\textcolor{blue}{.}\protect \\
Alicia Flórez Berdasco, María Elena de Cos Gómez, Jaime Laviada and
Fernando Las-Heras are with the Department of Electrical Engineering,
University of Oviedo, Spain. (e-mail: florezalicia@uniovi.es, medecos@uniovi.es;
laviadajaime@uniovi.es; flasheras@uniovi.es).}
}

\maketitle

\begin{abstract}
\textcolor{black}{An ultra-compact wearable antenna, for electronic
travel aid (ETA) applications in the 24.05-24.25\,GHz frequency band,
is presented. Th}e artificial magnetic conductor (AMC)-antenna combination
reduces the backward radiation to the wearing person, while improves
antenna's radiation properties and bandwidth without increasing its
area. Prototypes of the AMC-antenna have been fabricated and measured.
In order to test its performance for the application, imaging has
been conducted by means of synthetic aperture radar (SAR) techniques
by placing the antenna in the arm of a user to take advantage of natural
body movement. Electromagnetic images have been obtained and the target
has been identified, demonstrating the suitability of the AMC-antenna
for the ETA system.
\end{abstract}


\section{Introduction}

\IEEEPARstart{M}{ore} than 2 billion people worldwide suffer from
vision problems that, in the worst cases, limit their daily life \cite{ref1}.
The white cane is the most widespread mobility aid, along with guide
dogs. However, they are not enough for full autonomy, as the former
does not detect obstacles at torso height and the latter needs strong
training \cite{ref2}. Thus, numerous electronic travel aid systems
(ETA) have been developed in the last years. Most of them are based
on ultrasound sensors and video cameras \cite{ref2,ref3,ref4}, although
there are alternatives that use technologies such as NFC, infrared sensors, and others \cite{ref5}.

Radar technology stands out as one of the most promising technologies
for ETAs, especially at mmWave frequencies, where compact devices
suitable for portable applications are available. Moreover, it works
in all weather and visibility conditions and detects both, visible
and hidden objects. In addition, radars can be worn under clothing,
as the electromagnetic wave signals that they emit, can easily penetrate
through fabrics, making their use completely unobtrusive. Therefore,
due to all the advantages of radar technology, it seems entirely appropriate
as an alternative to the aforementioned systems.

The angular resolution of the system is limited by the radar aperture
\cite{ref6}, so synthetic aperture radar (SAR) techniques are going
to be implemented. This solution has been widely deployed for a large
number of applications in recent years such as nondestructive testing,
medical imaging or security \cite{ref8,ref9,ref10}. SAR is based
on moving the antenna and taking measurements at a set of positions
to coherently add all the reflected signals, in order to obtain high
resolution images. In this case, they are implemented by taking advantage
of the natural movement of the user \cite{ref10B}, due to the wearable
nature of the application.

Portable and wearable systems require small and compact antennas,
that can be easily carried by the user. Miniaturized, lightweight
and low-profile planar antennas are preferred for their comfort and
convenience for wearable applications. From the radar application
point of view, it is necessary to point out some considerations. Antennas
with narrow-beam radiation patterns, which enhance directivity, are
favored for long distance applications, so that a larger range of
operation is obtained. However, if near-field SAR techniques are going
to be explored, wide-beam radiation patterns seem better as the complete
scene can be illuminated at all the positions along the aperture.
Therefore, a trade-off solution between range and area of coverage
must be found. Moreover, our recent work evaluating the performance
of different antennas for ETA application illustrates how, for short
distance detection, antennas with low directivity exhibit better performance
than high gain antennas, which can provide anomalous results along
their electrically larger dimension \cite{ref10C}.

In this work, an AMC-backed twin Arrow Antenna designed, operative
in the 24.05-24.25\,GHz frequency band, for a wearable ETA system
is shown. In order to test its suitability for the target application,
SAR measurements taking advantage of the natural movement of the user
have been performed in the near-field of the aperture and electromagnetic
images have been obtained to detect targets.

The paper is organized as follows: first, the design procedure of
the antenna and the AMC are exposed. Later, the combination of both
structures is shown, and the results are compared with the original
antenna. Then, the fabrication methodology is explained, and the measurement
outcomes are confronted with the simulated ones. After that, the AMC-antenna
performance in the real radar application is shown and imaging results
are displayed and analyzed. Finally, conclusions are drawn.

\section{Antenna design}

\begin{figure}
\begin{centering}
\includegraphics[width=0.85\columnwidth]{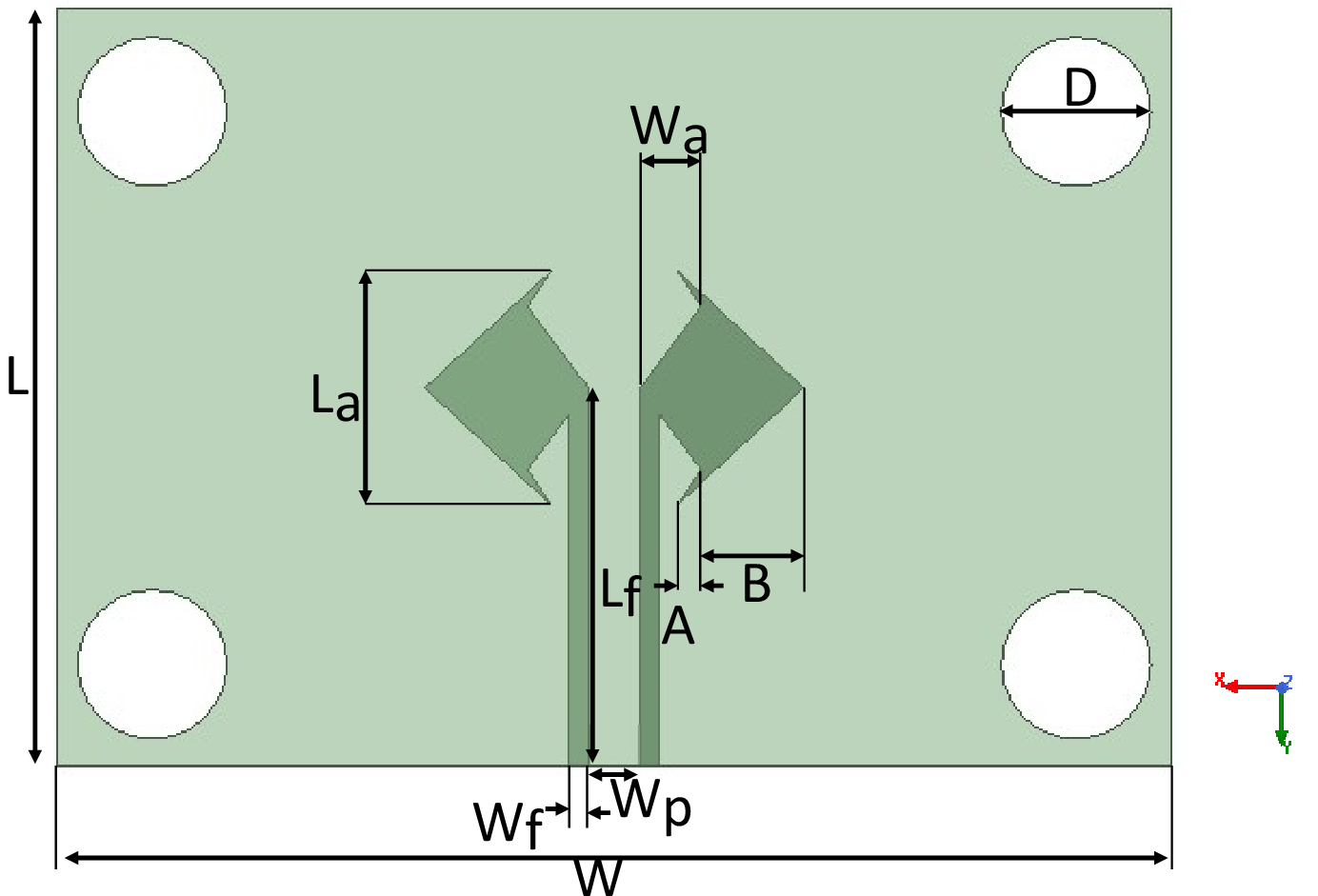}
\par\end{centering}
\centering{}\caption{Antenna geometry \label{fig:Antenna-geometry}}
\end{figure}

A novel twin arrow antenna, that operates in the 24.05 to 24.25\,GHz
frequency band, is presented. It follows the operating principles
of a planar dipole antenna, however, the shape of the radiating elements
has been modified to enhance its performance. RO3003 ($\varepsilon_{r}=3.0$,
$\tan\delta=0.0013$ and $h=0.762$\,mm) has been selected as dielectric
substrate since its intermediate value of relative dielectric permitivity
allows obtaining proper antenna size for fabrication suitability.
The two radiating elements and their narrow feeding lines are metallic
and have been optimized using the electromagnetic software simulator
HFSS. Fig. \ref{fig:Antenna-geometry} shows the antenna geometry,
and the final dimensions are summarized in Table \ref{tab:Antenna-dimension}.

\begin{table}
\begin{centering}
\caption{Antenna dimension (mm) \label{tab:Antenna-dimension}}
\par\end{centering}
\centering{}%
\begin{tabular}{|>{\centering}m{0.3cm}|>{\centering}m{0.3cm}|>{\centering}m{0.3cm}|>{\centering}m{0.3cm}|>{\centering}m{0.3cm}|>{\centering}m{0.3cm}|>{\centering}m{0.3cm}|>{\centering}m{0.3cm}|>{\centering}m{0.3cm}|>{\centering}m{0.3cm}|>{\centering}m{0.5cm}|}
\hline 
L & W & L\textsubscript{f} & W\textsubscript{f} & W\textsubscript{p} & L\textsubscript{a} & W\textsubscript{a} & A & B & D & h\tabularnewline
\hline 
9.6 & 14.1 & 4.8 & 0.25 & 0.45 & 0.87 & 2.96 & 0.3 & 1.6 & 1.9 & 0.762\tabularnewline
\hline 
\end{tabular}
\end{table}

The antenna exhibits proper impedance matching from 23.2 to 24.8\,GHz.
It provides 4.13\,dBi of gain (G) and 4.26~dB of directivity (D)
at 24.15\,GHz. Therefore, the radiation efficiency ($\eta$) is 97\,\%.
As the antenna is not backed with a ground plane, it shows a very
low front-to-back ratio (FTBR) of 0.25\,dB. All the radiation parameters
of the antenna are collected in Table \ref{tab:Radiation prop Antenna-Antenna+AMC}.

\subsection{Artificial Magnetic Conductor Design}

With the aim of reducing the backward radiation, an AMC metasurface
has been designed to back the antenna. Common metallic ground planes
(electric conductors) must be placed at a distance of at least one
quarter wavelength from the antenna. If this minimum distance is not
observed, there are no constructive interferences, and the antenna
is shorted. AMC is a type of metasurface that has the ability to reflect
the incident electromagnetic field in phase in the bandwidth that
comprises the frequencies for which the phase of the reflection coefficient
is between $\pm$90º \cite{ref11}. This characteristic allows placing
the AMC closer or even attached to the antenna \cite{ref11B,ref11C,ref11D,ref11E,ref11F}.

A squared metallization has been situated over a grounded dielectric
(RO3003) to create the AMC unit-cell. A simple geometry has been selected
to facilitate its fabrication. The geometry of the unit-cell is shown
in Fig. \ref{fig:Antenna S11 and AMC refelction phase} with its reflection
coefficient. Floquet port and periodic boundary conditions were considered
in simulation for the unit-cell optimization based on the phase of
the reflection coefficient. The resulting dimensions of the unit-cell
are $P=3.2$\,mm, $g=0.5$\,mm and $h=0.762$\,mm. The resonance
frequency of the AMC is $26.6$\,GHz and the AMC operating bandwidth
is 22-31.1\,GHz (see Fig. \ref{fig:Antenna S11 and AMC refelction phase}).

\subsection{AMC-antenna Combination}

\begin{figure}
\centering{}\includegraphics[width=1\columnwidth]{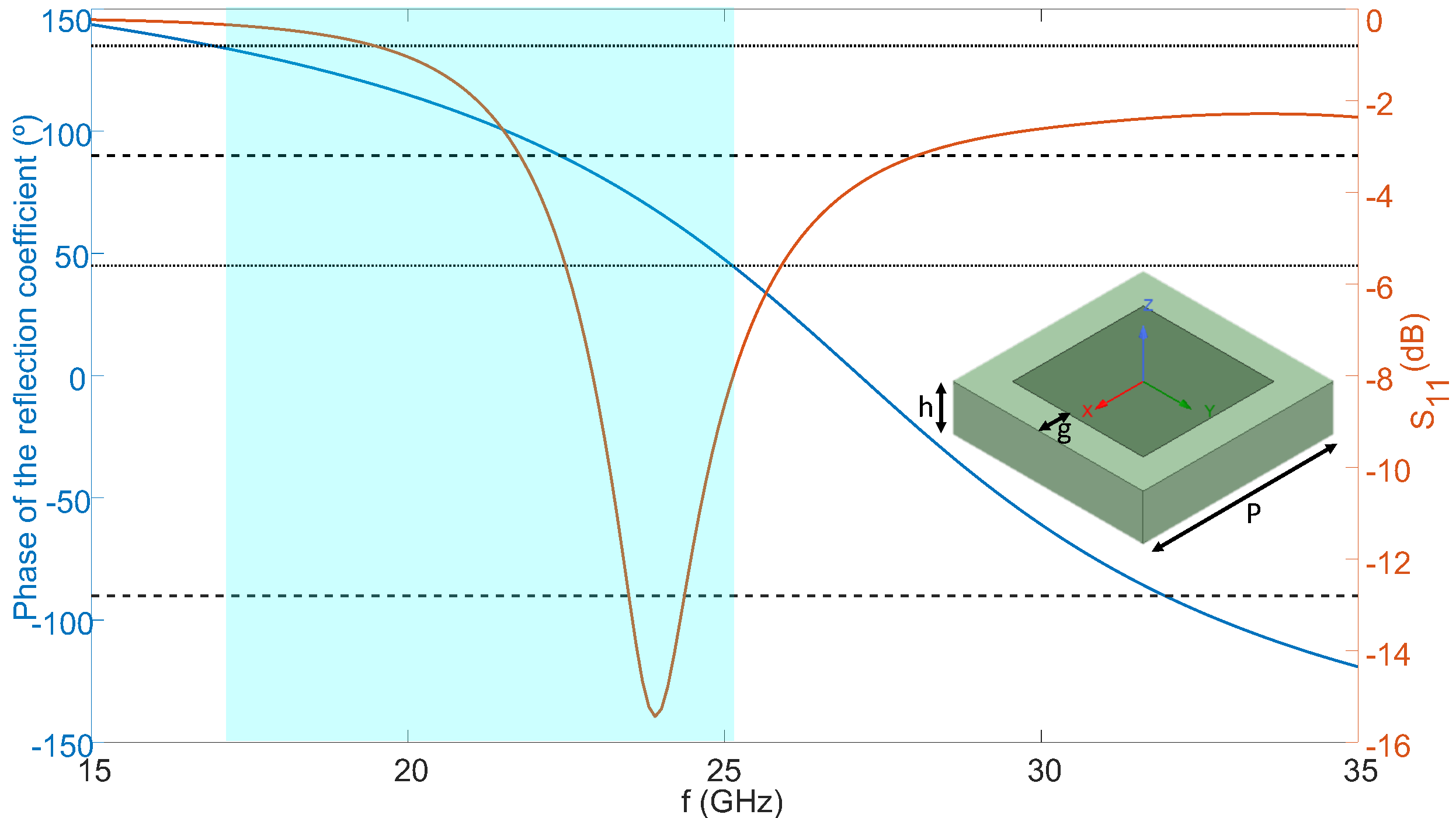}\caption{Simulated antenna reflection coefficient and AMC reflection phase
and its geometry. Light blue strip shows the working band of the AMC.\label{fig:Antenna S11 and AMC refelction phase}}
\end{figure}

The AMC metasurface has been arranged under the antenna, so that the
backward radiation to the wearing person will be reduced. The $90\text{º}\pm45\text{º}$
reflection phase region of the AMC, is the optimum operating bandwidth
when the metasurface is combined with a dipole antenna \cite{ref12}.
Fig. \ref{fig:Antenna S11 and AMC refelction phase} shows the antenna
reflection coefficient with the AMC reflection phase. It can be observed
that the resonance frequency of the antenna is included in the frequency
band in which the reflection phase of the AMC is between $90\text{º}\pm45\text{º}$,
which has been highlighted in light blue.

\begin{figure}
\centering{}\includegraphics[width=1\columnwidth]{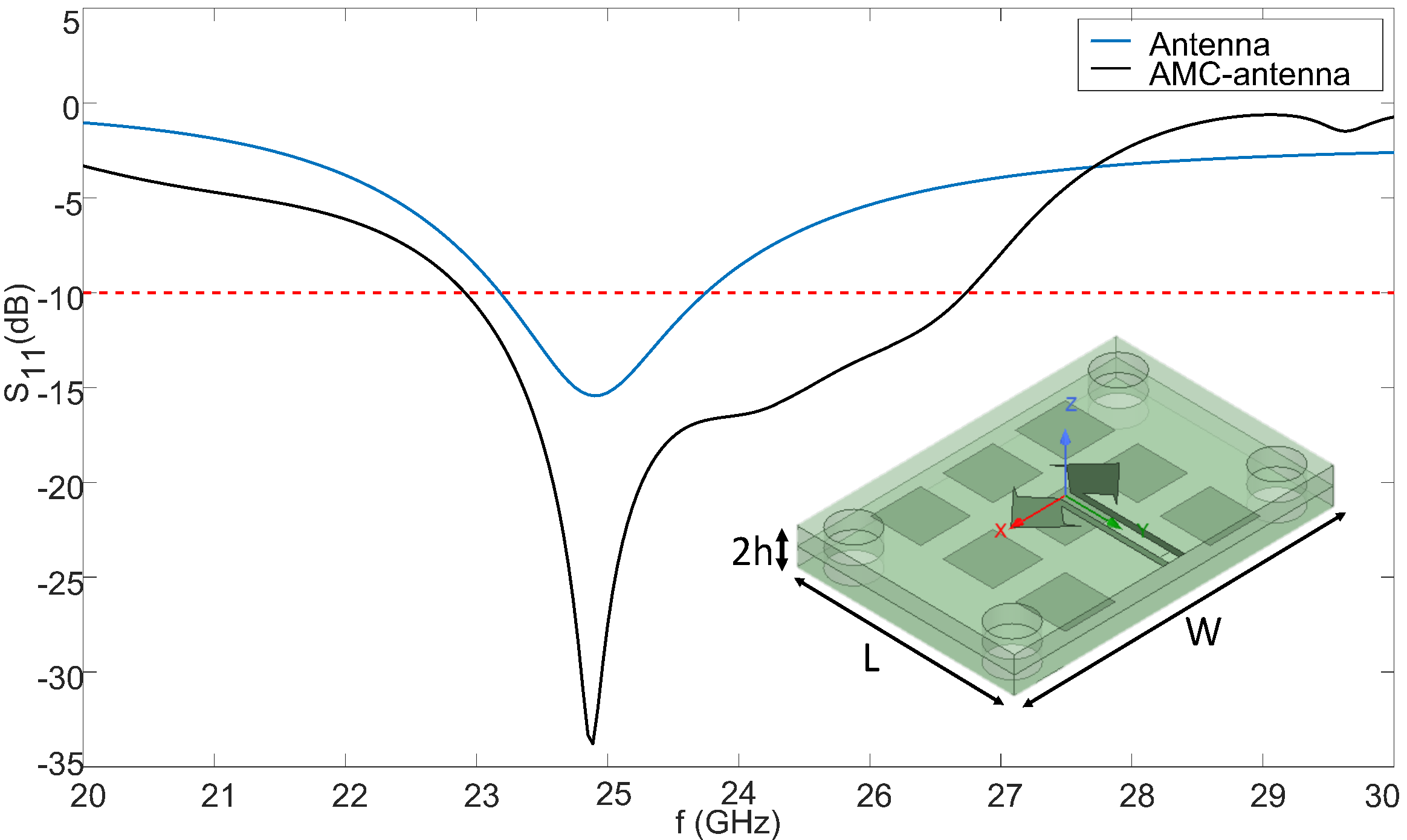}\caption{$S_{11}$ comparison for the antenna and the AMC-antenna. \label{fig:S11 antenna_antenna+AMC}}
\end{figure}

\begin{figure}
\includegraphics[width=1\columnwidth]{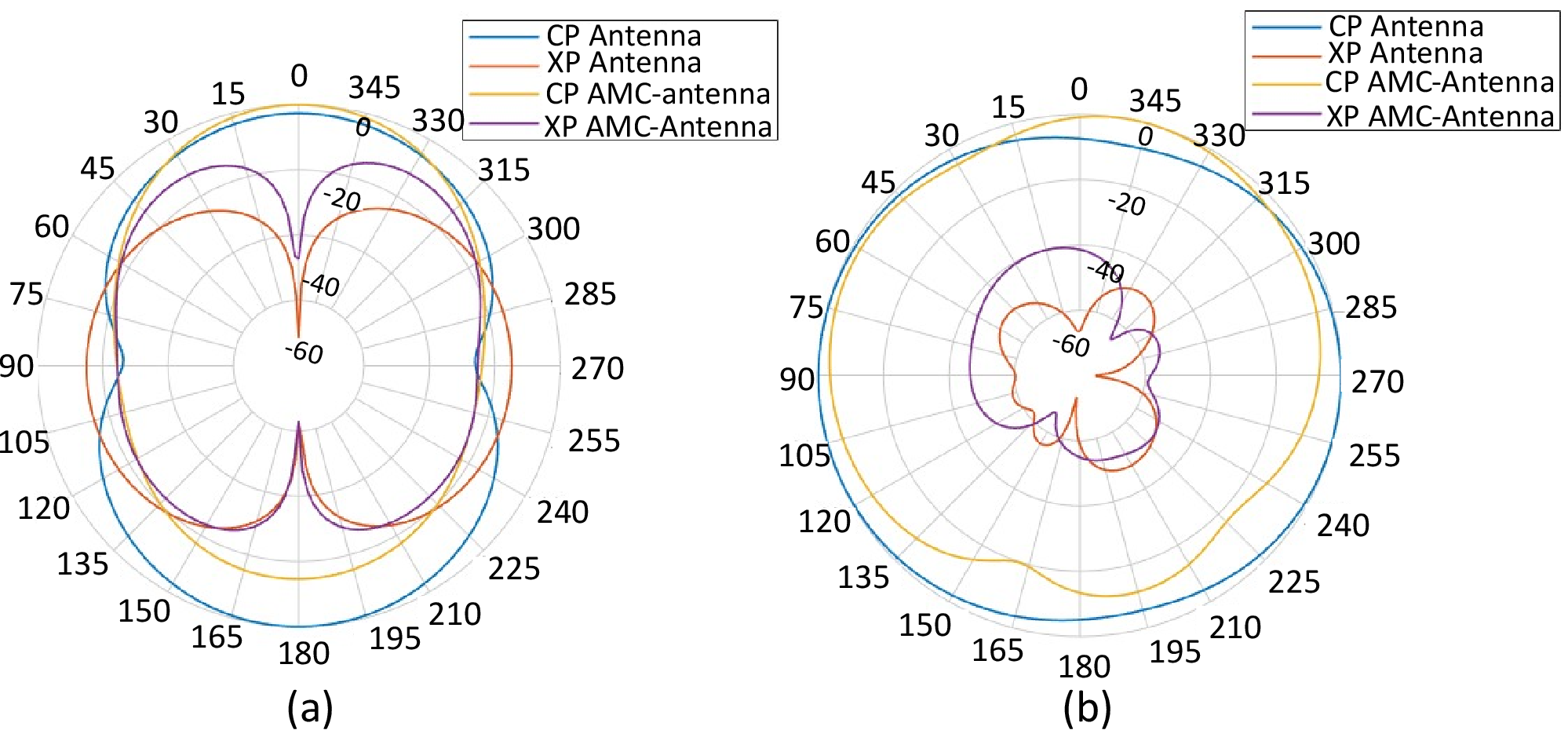}\caption{Radiation pattern cuts. (a) Phi=0º, (b) Phi=90º. \label{fig:Radiation patterns}}
\end{figure}

The AMC has been placed under the antenna without any gap between
them. It should be noted that the unit-cell under the feed lines has
been removed so as not to disturb the feeding network. The final geometry
of the combined structure is presented in Fig. \ref{fig:S11 antenna_antenna+AMC},
with the reflection coefficient of both, the antenna alone and the
AMC-antenna. The AMC-antenna shows a much wider bandwidth than the
antenna without metasurface, however, both operate in the target frequency
band. The radiation pattern cuts are depicted in Fig. \ref{fig:Radiation patterns}.
It can be observed that the AMC metasurface reduces the backward radiation
and slightly modifies the radiation pattern, so that a more directive
one is observed. Table \ref{tab:Radiation prop Antenna-Antenna+AMC}
collects the main radiation properties of both antennas. It can be
concluded that the AMC-antenna improves around 2\,dB both, directivity
and gain with respect to the original antenna. The radiation efficiency
keeps high across the whole bandwidth. The great advantage of placing
an AMC metasurface behind the antenna is raising the FTBR parameter
by approximately 15\,dB, so that, when the AMC is placed under the
antenna, the backward radiation to the body is significantly reduce,
which is essential in wearable devices. Therefore, all the radiation
properties and the bandwidth of the original antenna have been improved
while preserving the initial area. The unavoidable thickness increase
(only due to the AMC as there is no layer between it and the antenna)
however has no impact on the resulting antenna comfort and wearability.
Hence, an ultra-compact and operative antenna has been achieved to
be used in ETA applications.

\begin{table}
\begin{centering}
\caption{Radiation properties for the antenna and AMC-antenna. \label{tab:Radiation prop Antenna-Antenna+AMC}}
\par\end{centering}
\centering{}%
\begin{tabular}{|>{\raggedright}m{0.8cm}|>{\centering}m{0.7cm}||>{\centering}p{0.95cm}|>{\centering}p{0.85cm}|>{\centering}p{0.8cm}|>{\centering}p{0.65cm}|>{\centering}p{0.9cm}|}
\cline{2-7} \cline{3-7} \cline{4-7} \cline{5-7} \cline{6-7} \cline{7-7} 
\multicolumn{1}{>{\raggedright}m{0.8cm}|}{} & BW(\%) & F (GHz) & G (dBi) & D (dB) & $\eta$ (\%) & FTBR (dB)\tabularnewline
\hline 
\multirow{3}{0.8cm}{Antenna} & \multirow{3}{0.7cm}{6.21} & 24.05 & 4.14 & 4.26 & 97.3 & 0.25\tabularnewline
\cline{3-7} \cline{4-7} \cline{5-7} \cline{6-7} \cline{7-7} 
 &  & 24.15 & 4.13 & 4.26 & 97.1 & 0.25\tabularnewline
\cline{3-7} \cline{4-7} \cline{5-7} \cline{6-7} \cline{7-7} 
 &  & 24.25 & 4.01 & 4.26 & 94.4 & 0.25\tabularnewline
\hline 
\multirow{3}{0.8cm}{AMC- antenna} & \multirow{3}{0.7cm}{15.7} & 24.05 & 6.72 & 6.72 & 100 & 14.32\tabularnewline
\cline{3-7} \cline{4-7} \cline{5-7} \cline{6-7} \cline{7-7} 
 &  & 24.15 & 6.73 & 6.74 & 99.8 & 15.58\tabularnewline
\cline{3-7} \cline{4-7} \cline{5-7} \cline{6-7} \cline{7-7} 
 &  & 24.25 & 6.74 & 6.77 & 99.3 & 16.12\tabularnewline
\hline 
\end{tabular}
\end{table}

\subsection{Fabrication and Measurement}

\begin{figure}
\begin{centering}
\includegraphics[width=1\columnwidth]{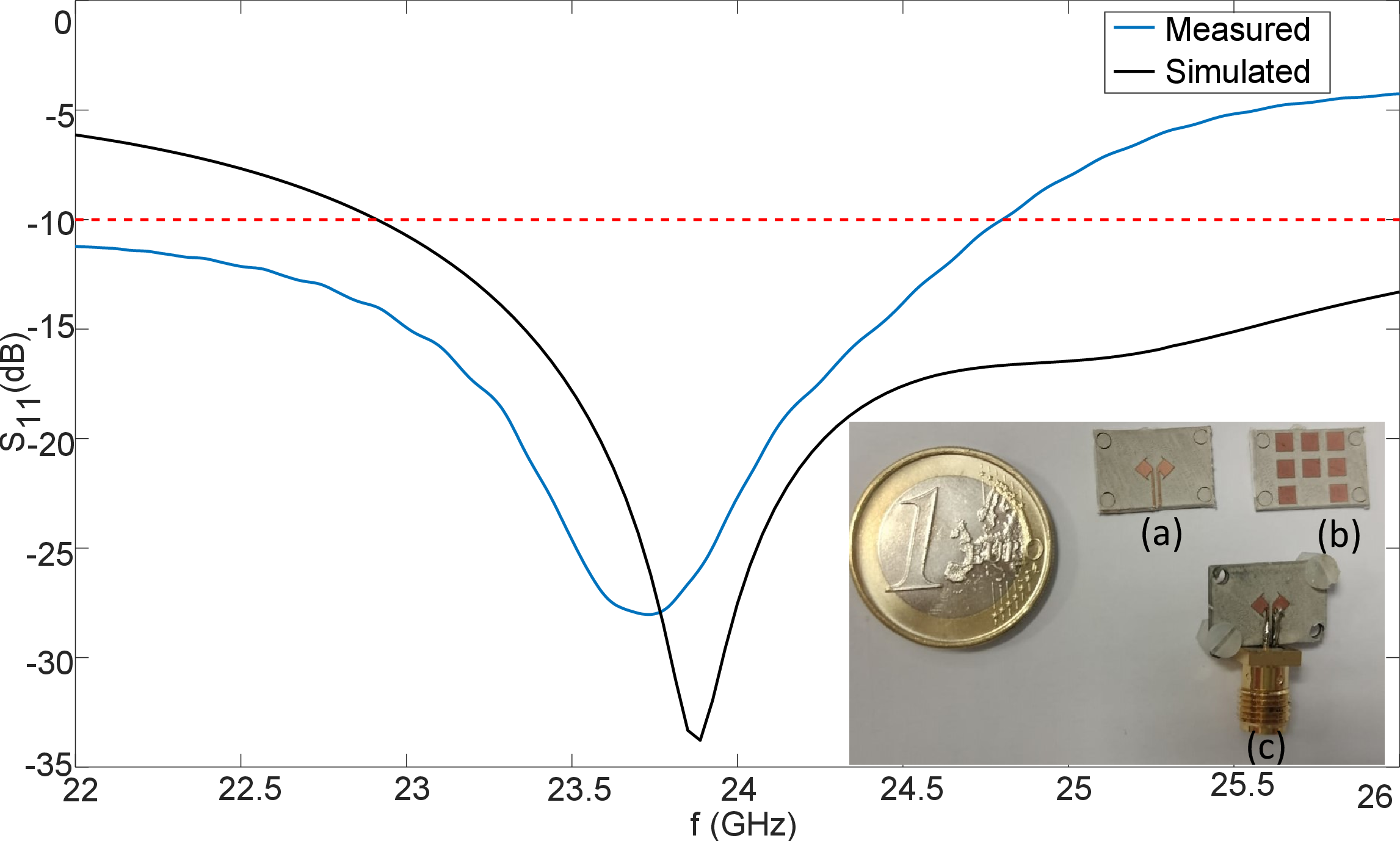}
\par\end{centering}
\centering{}\caption{Comparison between the measured and simulated $S_{11}$ of the AMC-antenna
together with the antennas prototype. (a) Antenna prototype, (b) AMC
prototype, (c) AMC-antenna prototype.\label{fig:Measured prototypes}}
\end{figure}

Prototypes of the AMC-antenna have been fabricated using laser micromachining.
Each element (antenna and AMC) is manufactured separately and then,
they have been fixed using nylon screws, as it can be appreciated
in Fig. \ref{fig:Measured prototypes}. It depicts measured and simulated
reflection coefficient results of the AMC-antenna. The measured trace
has been slightly shifted downwards in frequency; however, it shows
proper impedance matching in the frequency bandwidth of interest.
Discrepancies between simulated and measured results can be attributed
to several issues. On the one hand, manufacturing tolerances can cause
the displacement of the resonance frequency. On the other hand, the
connector is soldered by hand to the narrow strips and its size is
very large compared to the antenna, which can produce these mismatches.
In addition, the antenna and the AMC are aligned through the screws,
but it is possible that small misalignment may occur, which can result
in frequency shifts. Nevertheless, the measured trace has deviated
by just 0.84\% from the simulated one.

\subsection{\textcolor{black}{Comparison with Other Millimeter-Wave Antennas
at 24 GHz}}

A brief comparison with the state-of-the-art antennas at 24 GHz is
provided to endorse the achievements of this work. Despite using dielectric
with practically the same $\varepsilon_{r}$, the AMC-antenna outperforms
\cite{ref14,ref15} in area and bandwidth. It also overcomes \cite{ref14}
in gain. Although it provides slightly lower gain than \cite{ref15},
the AMC-antenna triples its radiation efficiency. Moreover, \cite{ref15}
is fabricated in paper, which is not the best material for a portable
systems, as it can be easily damaged. Compared to those on substrates
with slightly higher $\varepsilon_{r}$, the AMC-antenna overcomes
\cite{ref15B,ref15C} in bandwidth and gain, with a more compact design
in area. In addition, it exhibits better FTBR than \cite{ref15C}.
\cite{ref16} presents a bigger antenna in comparison with the one
of this paper, even though a material with a higher $\varepsilon_{r}$
is used. Besides, it provides half the bandwidth, despite requiring
a greater thickness of material. It shows better gain, however, $\eta$
and FTBR are not given.

\begin{table}[H]
\centering{}\caption{State-of-the-art antennas at 24 GHz.}
\begin{tabular}{|>{\centering}p{1.5cm}|>{\centering}m{1.8cm}|>{\centering}m{0.5cm}|>{\centering}m{0.5cm}|>{\centering}m{0.5cm}|>{\centering}m{0.5cm}|>{\centering}m{0.6cm}|}
\cline{2-7} \cline{3-7} \cline{4-7} \cline{5-7} \cline{6-7} \cline{7-7} 
\multicolumn{1}{>{\centering}p{1.5cm}|}{} & Size $(\text{mm}^{3})$ & $\varepsilon_{r}$ & BW (\%) & G (dBi) & $\eta$ (\%) & FTBR (dB)\tabularnewline
\hline 
\cite{ref14} & $76\times76\times0.13$ & 3 & 3.8 & 4.8 & - & -\tabularnewline
\hline 
\cite{ref15} & $20\times25\times0.23$ & 2.9 & 2.3 & 7.4 & 35 & -\tabularnewline
\hline 
\cite{ref15B} & $36.5\times53\times0.1$ & 3.35 & 0.8 & 5.81 & - & -\tabularnewline
\hline 
\cite{ref15C} & $23\times15\times0.254$ & 3.48 & 6.6 & 4.24 & - & 9.23\tabularnewline
\hline 
\cite{ref16} & $24\times24\times1.6$ & 6.4 & 8 & 9 & - & -\tabularnewline
\hline 
AMC-antenna & $9.6\times14.1\times1.524$ & 3 & 15.7 & 6.73 & 99.8 & 15.58\tabularnewline
\hline 
\end{tabular}
\end{table}

\section{Imaging testing}

Antenna performance is evaluated experimentally for the intended ETA
application. For this purpose, SAR techniques for high-resolution
electromagnetic imaging are implemented to map the nearby environment.
This is achieved by taking advantage of the natural movement of the
body, so a resolution better than the one obtained with a single position,
is reached \cite{ref10B}.

Electromagnetic images are computed by means of a flexible sum-and-delay
algorithm for monostatic acquisition \cite{ref16B}:

\begin{equation}
\hat{\rho}(\overrightarrow{r}')={\displaystyle \sum_{m=1}^{M}\sum_{n=1}^{N}S(m,n)\cdot e^{j2k_{m}\left|\overrightarrow{r}'-\overrightarrow{r}_{n}\right|}},
\end{equation}
in which $\rho\left(\overrightarrow{r}'\right)$ stands for the reflectivity
of the target, $\overrightarrow{r}'$ represents to the pixel where
reflectivity will be calculated, $S$ depicts the acquired data corresponding
to the $s_{11}$ parameter after subtracting the background and calibrating
the phase shift, $M$ indicates the number of frequencies and $N$
refers to the acquisition points, $k_{m}$ refers to the wavenumber
at the $m$-th frequency and $\overrightarrow{r}_{n}$ appoints the
$n$-th measured position.

\begin{figure}
\subfloat[]{\includegraphics[width=0.5\columnwidth]{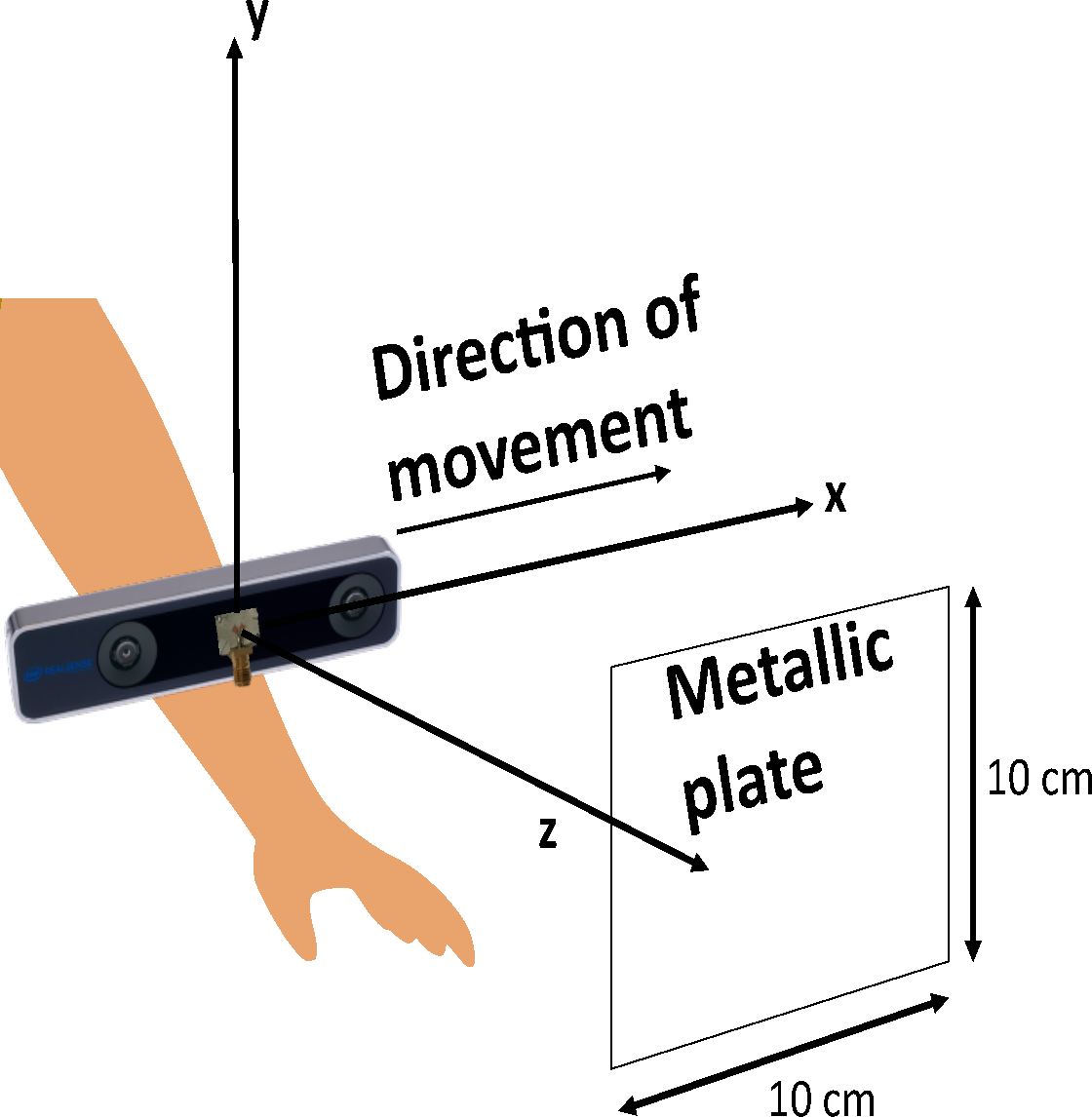}

}\subfloat[]{\includegraphics[width=0.5\columnwidth]{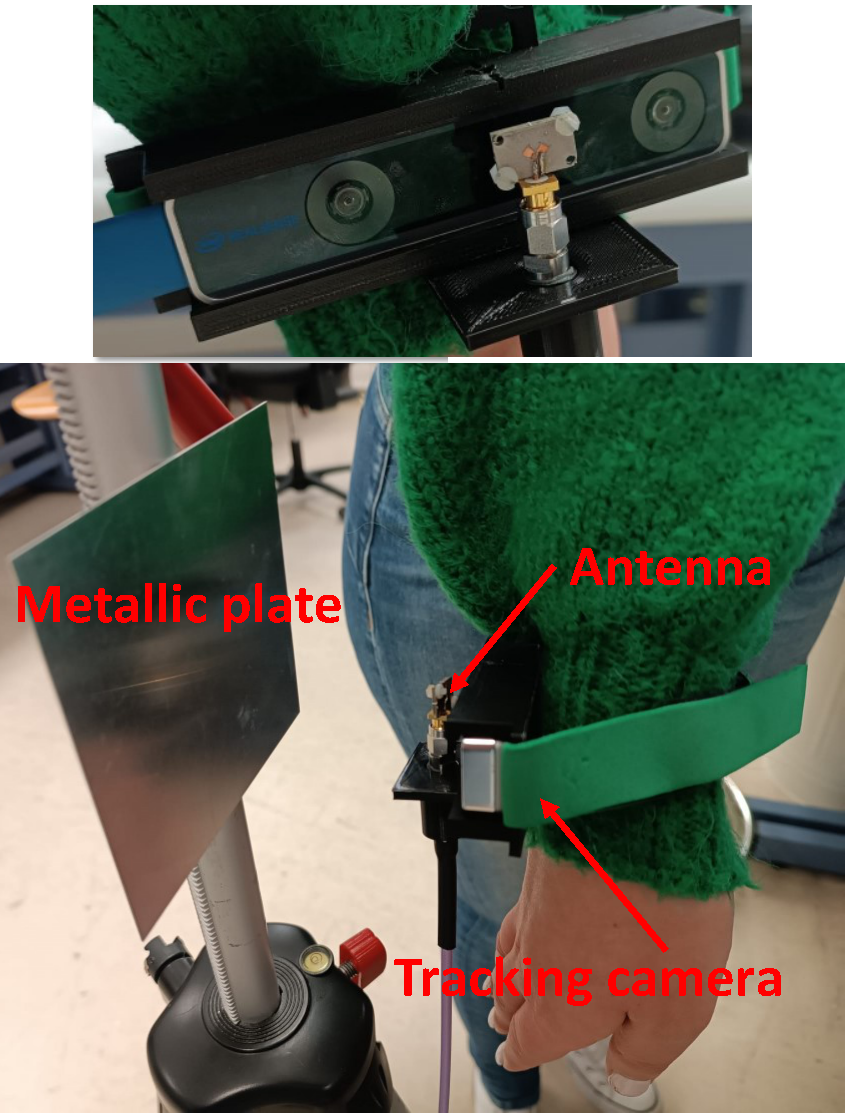}}\caption{Measurement set-up. (a) Schematic, (b) Real laboratory set-up.\label{fig:MeasurementSetup}}
\end{figure}

Fig \ref{fig:MeasurementSetup}(a) depicted the pure monostatic set-up
used to perform the measurements. A metallic plate of $10\times10$\,cm
has been used as a target. The movement of the antenna is recorded
using a tracking camera. In particular, the Intel® RealSense™ Tracking
Camera T265 \cite{ref17} is used to register the measurement positions.
The antenna and the camera have been placed in a 3D printed gadget,
in order to move them together, as well as to fix them to the user's
arm, as can be observed in Fig \ref{fig:MeasurementSetup}(b).

The antenna is moved by means of simple arm swings. The measurement
point interval was cropped to $12$\,cm from the target as the tracking
device accumulate positioning errors and only positions within a short
period are consistent. The target was placed at $10\,\text{cm}$.
The scattering parameter $S$ is measured with a vector network analyzer
(VNA), in order to emulate a stepped-frequency continuous-wave (SFCW)
radar. Table \ref{tab:SAR Param} summarizes the SAR configuration
considerations set up to perform the measurement. It is necessary
to clarify that the measurement has been conducted in an extended
frequency band from 22 to 26 GHz, to obtain a good range resolution
that allows to distinguish the target and its range position. The
corresponding measurement dataset, can be found in \cite{ref17B}.

\begin{table}
\caption{Experimental parameters for the SAR measurement.\label{tab:SAR Param}}
\begin{tabular}{|>{\centering}p{13mm}|>{\centering}p{13mm}|>{\centering}p{13mm}|>{\centering}p{13mm}|>{\centering}p{13mm}|}
\hline 
Centre frequency (GHz) & Bandwidth (GHz) & Number of frequencies & Synthetic aperture length (cm) & Target dimensions (cm)\tabularnewline
\hline 
24 & 4 & 3201 & 12 & $10\times10$\tabularnewline
\hline 
\end{tabular}
\end{table}

\begin{figure}
\centering{}\includegraphics[width=0.8\columnwidth]{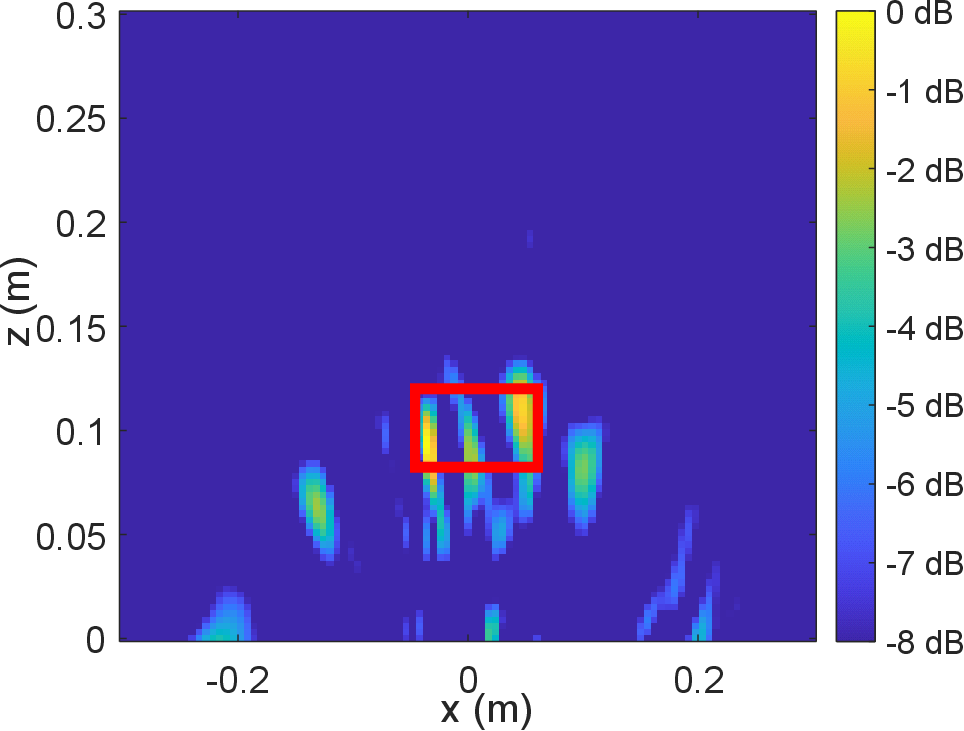}\caption{Electromagnetic image obtained with the AMC-antenna.\label{fig:ElectromagneticImage}}
\end{figure}

Fig. \ref{fig:ElectromagneticImage} presents the electromagnetic
image obtained with the AMC-antenna. Although there are artifacts
due to the irregular sampling \cite{key-1}, the target is correctly
detected at $10$\,cm from the radar, that has been highlighted in
red color in the image, which agrees with its real position. Besides, it
can be observed that the target width ($x$-axis) matches the actual
object size of $10$\,cm, as expected, since the resolution is only
along the $x$-axis. Therefore, the antenna performance is suitable
for the pursued application.

\section{Conclusions}

A novel ultra-compact and lightweight AMC-backed twin arrow wearable
antenna, suitable for ETA applications, has been achieved. The FTBR
and radiation properties of the initial antenna have been improved
by backing it with an AMC. Significant improvement has been obtained
in the FTBR, which is key for the intended application, in order to
avoid user radiation. In addition, the radiation parameters and the
operation bandwidth of the antenna have been enhanced, without increasing
its area.

Experimental measurements for electronic travel aid purposes were
conducted by generating electromagnetic images of the surrounding
environment. These images were generated by taking advantage of the
natural body movement and exploiting SAR techniques. It was shown
how the antenna can take advantage of those techniques to improve
the resolution along the movement direction, enabling a better detection
of the targets.


\begin{thebibliography}{10}
\bibitem{ref1} World Health Organization, January 2024. {[}Online{]}
Available: https://www.who.int/news-room/fact-sheets/detail/blindness-and-visual-impairment.

\bibitem{ref2} D. Abreu, J. Toledo, B. Codina, and A. Suárez, “Low-cost
ultrasonic range improvements for an assistive device,” Sensors, vol.
21, no. 12, p. 4250, Jun. 2021, doi: 10.3390/s21124250.

\bibitem{ref3} C. Gearhart, A. Herold, B. Self, C. Birdsong and L.
Slivovsky, \textquotedbl Use of ultrasonic sensors in the development
of an Electronic Travel Aid,\textquotedbl{} 2009 IEEE Sensors Applications
Symposium, New Orleans, LA, USA, 2009, pp. 275-280, doi: 10.1109/SAS.2009.4801815.

\bibitem{ref4}A. S. Rao, J. Gubbi, M. Palaniswami and E. Wong, \textquotedbl A
vision-based system to detect potholes and uneven surfaces for assisting
blind people,\textquotedbl{} 2016 IEEE International Conference on
Communications (ICC), Kuala Lumpur, Malaysia, 2016, pp. 1-6, doi:
10.1109/ICC.2016.7510832.

\bibitem{ref5}E. Cardillo and A. Caddemi, “Insight on electronic
travel aids for visually impaired people: A review on the electromagnetic
technology,” Electronics, vol. 8, no. 11, p. 1281, Nov. 2019, doi:
10.3390/electronics8111281.

\bibitem{ref6} I. M. Skolnik, Introduction to RADAR Systems. New
York, NY, USA: McGraw-Hill, 1980

\bibitem{ref8} D. M. Sheen, D. L. McMakin, and T. E. Hall, “Three-dimensional
millimeter-wave imaging for concealed weapon detection,” IEEE Trans.
Microw. Theory Techn., vol. 49, no. 9, pp. 1581--1592, Sep. 2001,
doi: 10.1109/22.942570

\bibitem{ref9} S. Agarwal, B. Kumar and D. Singh, \textquotedbl Non-invasive
concealed weapon detection and identification using V band millimeter
wave imaging radar system,\textquotedbl{} 2015 National Conference
on Recent Advances in Electronics \& Computer Engineering (RAECE),
Roorkee, India, 2015, pp. 258-262, doi: 10.1109/RAECE.2015.7510202.

\bibitem{ref10}M. T. Bevacqua, S. Di Meo, L. Crocco, T. Isernia and
M. Pasian, \textquotedbl Millimeter-Waves Breast Cancer Imaging via
Inverse Scattering Techniques,\textquotedbl{} in IEEE Journal of Electromagnetics,
RF and Microwaves in Medicine and Biology, vol. 5, no. 3, pp. 246-253,
Sept. 2021, doi: 10.1109/JERM.2021.3052096.

\bibitem{ref10B}H. F. Álvarez, G. Álvarez-Narciandi, F. Las-Heras
and J. Laviada, “System Based on Compact mmWave Radar and Natural
Body Movement for Assisting Visually Impaired People,” in IEEE Access,
vol. 9, pp. 125042-125051, 2021, doi: 10.1109/ACCESS.2021.3110582.

\bibitem{ref10C}A. F. Berdasco, J. Laviada, M. E. de Cos Gómez and
F. Las-Heras, \textquotedbl Performance Evaluation of Millimeter-Wave
Wearable Antennas for Electronic Travel Aid,\textquotedbl{} in IEEE
Transactions on Instrumentation and Measurement, vol. 72, pp. 1-10,
2023, Art no. 4507510, doi: 10.1109/TIM.2023.3320736.

\bibitem{ref11}A. Flórez Berdasco, M.E. de Cos Gómez, H. Fernández
Álvareza nd F. Las-Heras, ``Millimeter wave array-HIS antenna for
imaging applications,'' in Appl. Phys. A 129, 397 (2023), doi: 10.1007/s00339-023-06676-0.

\bibitem{ref11B} M. Mantash, A.-C. Tarot, S. Collardey, and K. Mahdjoubi,
“Design methodology for wearable antenna on artificial magnetic conductor
using stretch conductive fabric,” Electron. Lett., vol. 52, no. 2,
pp. 95--96, Jan. 2016.

\bibitem{ref11C}A. Alemaryeen and S. Noghanian, \textquotedbl On-Body
Low-Profile Textile Antenna With Artificial Magnetic Conductor,\textquotedbl{}
in IEEE Transactions on Antennas and Propagation, vol. 67, no. 6,
pp. 3649-3656, June 2019, doi: 10.1109/TAP.2019.2902632.

\bibitem{ref11D}Ayd R. Saad, A., Hassan, W.M. \& Ibrahim, A.A. A
monopole antenna with cotton fabric material for wearable applications.
Sci Rep 13, 7315 (2023). doi: 10.1038/s41598-023-34394-3

\bibitem{ref11E}K Srilatha et al 2021 J. Phys.: Conf. Ser. 1804 012189

\bibitem{ref11F}Saha, P. \& Mitra, D. \& Parui, S.K.. (2021). Control
of Gain and SAR for Wearable Antenna Using AMC Structure. Radioengineering,
vol. 30, pp. 81-88. doi: 10.13164/re.2021.0081.

\bibitem{ref12}Fan Yang and Y. Rahmat-Samii, \textquotedbl Reflection
phase characterizations of the EBG ground plane for low profile wire
antenna applications,\textquotedbl{} in IEEE Transactions on Antennas
and Propagation, vol. 51, no. 10, pp. 2691-2703, Oct. 2003, doi: 10.1109/TAP.2003.817559.

\bibitem{ref14}J. Bito, V. Palazzi, J. Hester, R. Bahr, F. Alimenti,
P. Mezzanotte, L. Roselli, M. M. Tentzeris, “Millimeter-wave ink-jet
printed RF energy harvester for next generation flexible electronics,”
2017 IEEE Wireless Power Transfer Conference (WPTC), Taipei, Taiwan,
10--12 May 2017; pp. 1--4.

\bibitem{ref15}P. Mezzanotte, C. Mariotti, M. Virili, M. Poggiani,
G. Orecchini, F. Alimenti and L. Roselli, “24-GHz Patch antenna array
on cellulosebased materials for green wireless internet applications,”
IET Science, Measurement \& Technology,vol. 8 pp. 342-349, 2014

\bibitem{ref15B}N. Kathuria, B.-C. Seet, “24 GHz Flexible Antenna
for Doppler Radar-Based Human Vital Signs Monitoring,” Sensors 2021,
21, 3737.

\bibitem{ref15C}Park S, Kim S, Kim DK, Choi J, Jung K-Y. ``Numerical
Study on the Feasibility of a 24 GHz ISM-Band Doppler Radar Antenna
for Near-Field Sensing of Human Respiration in Electromagnetic Aspects,''
Applied Sciences. 2020; 10(18):6159, doi: 10.3390/app10186159.

\bibitem{ref16}F. A. Ghaffar, M. U. Khalid, K. N. Salama and A. Shamim,
“24-GHz LTCC Fractal Antenna Array SoP With Integrated Fresnel Lens,”
IEEE Antennas and Wireless Propagation Letters, vol. 10, pp. 705-708,
2011.

\bibitem{ref16B}M. Soumekh, Synthetic Aperture Radar Signal Processing,
New York, NY, USA: Wiley, vol. 7, 1999.

\bibitem{ref17}Intel® RealSense™ Tracking Camera T265, January 2024.
{[}Online{]} Available: https://www.intelrealsense.com/visual-inertial-tracking-case-study/.

\bibitem{ref17B}Alicia Florez Berdasco, María Elena de Cos Gómez
, Jaime Laviada, Fernando Las-Heras, February 1, 2024, \textquotedbl Experimental
SAR measurements for electronic travel aid purposes\textquotedbl ,
IEEE Dataport, doi: https://dx.doi.org/10.21227/dcdt-1e95.

\bibitem{key-1}J. Laviada, G. Álvarez-Narciandi and F. Las-Heras,
\textquotedbl Artifact Mitigation for High-Resolution Near-Field
SAR Images by Means of Conditional Generative Adversarial Networks,\textquotedbl{}
in IEEE Transactions on Instrumentation and Measurement, vol. 71,
pp. 1-11, 2022, Art no. 8006011, doi: 10.1109/TIM.2022.3200107
\end{thebibliography}
\end{document}